\documentclass[aip,reprint,amssymb,twocolumn]{revtex4}  

\usepackage{dcolumn}
\usepackage{bm}

\usepackage[utf8]{inputenc}
\usepackage[T1]{fontenc}
\usepackage{mathptmx}
\usepackage{etoolbox}

\usepackage{amsmath}%
\usepackage{graphicx}%
\usepackage{color}%
\usepackage{lineno,hyperref}
\usepackage{bm}

\newcommand{\ham}{$\hat{H}$}

\newcommand{\hamt}{$\hat{H}(t)$}

\newcommand{\kai}{$\chi$}
\newcommand{\kaiinv}{$\chi^{-1}$}

\makeatletter
\def\@email#1#2{%
 \endgroup
 \patchcmd{\titleblock@produce}
  {\frontmatter@RRAPformat}
  {\frontmatter@RRAPformat{\produce@RRAP{*#1\href{mailto:#2}{#2}}}\frontmatter@RRAPformat}
  {}{}
}%
\makeatother
\begin{document}
\title{Optical excitation from anti-causally corrected real-time dynamics in a minimal basis}
\author{Joydev De$^{1,2}$, Manoar Hossain$^{1,3}$, Joydeep Bhattacharjee$^{1}$ \\
\small{\textit{$^1$National Institute of Science Education and Research,\\ 
A CI of Homi Bhabha National Institute, Jatani, Odisha - 752050, India\\
$^2$Sahid Kshudiram Smriti Mahavidyalaya, Gangaramchak, Paschim Medinipur, West Bengal, India.\\
$^3$Institut f{\"{u}}r Physik and IRIS Adlershof, Humboldt-Universit{\"{a}}t zu Berlin, Zum Gro\ss en Windkanal 2, D-12489 Berlin, Germany}}}

\begin{abstract}
Here we demonstrate 
workably accurate estimation of optical excitation threshold for large systems comprising of 
hundreds of atoms through an anti-causally corrected(ACC) real-time dynamics(RTD) approach 
implemented in a minimal tight-binding basis constituted by the directed hybrid atomic Wannier orbitals. 
A correction to the Hamiltonian is applied anti-causally at all time steps to account for 
electron-hole interaction using the density-density response function. 
Minimality of basis and ease of transferability of parameters to large systems 
arises from the directed nature of the Wannierized hybrid basis orbitals used.
With self-energy corrected TB parameters evaluated at the $DFT+G_0W_0$ level, the proposed ACC-RTD scheme can be 
systematically parametrized to render optical excitation threshold
for systems of experimentally realizable length-scales through inexpensive computation.
\end{abstract}
\maketitle
\section{Introduction}

%
Over the last few decades a key motivation to study 
nano-structures of a different size and shape \cite{van1998changes,brus1984electron}
primarily of elemental or compound semi-conducting materials, 
has been to manipulate light-matter interaction aimed at applications ranging from controlled absorption, emission and 
harvesting of light \cite{bostedt2004strong,alivisatos1996perspectives}. 
 In this direction elements of the $p$-block having valence electrons in $2p,3p$ and $4p$ orbitals, have been in the forefront 
\cite{landt2009optical}. 
%
Effective theoretical prediction of new device possibilities in this direction thus crucially depends on accurate estimation of
optical-gap of nano-structures \cite{raty2005first,demjan2014electronic,yin2014quasiparticle,tiago2006optical}. 
 However, they still consists of few hundreds to thousands of atoms which makes it a considerable computational challenge 
to compute optical excitations accurately.
%
%

Time dependent density functional theory(TDDFT) using the LDA exchange-correlation (xc) has been traditionally used to 
compute optical excitation in extended systems \cite{raty2005first,demjan2014electronic}. 
%
However as system size shrinks, a more structured correlation among electrons 
causes loss of accuracy of the ground-state calculation and also the description of electron-hole interaction with static local xc functionals used in DFT. 
%
While the ground-state energetics can be improved using 
hybrid pseudo-potentials \cite{heyd2003hybrid}, 
the most accurate computation of the energetics of the ground-state 
starting with the KS single particle levels, is obtained perturbatively as their self-energy corrections 
due to a non-local and dynamic self-energy operator derived within the GW approximation of the many-body perturbation
theory 
\cite{hedin1965new,hedin1971explicit,hybertsen1986electron}. 
DFT+GW computation however is expensive due to convergence of the dielectric function and the 
self-energy operator, with respect to unoccupied states, typically requiring in thousands.
Subsequently, TDLDA calculation with scissor correction as per the shift in single particle levels
as per their self-energy corrections, has been shown to render correct optical gap for
smaller carbon nano-diamonds \cite{demjan2014electronic}.
Scissor correction has also been extended to the real time TDDFT\cite{wang2019real} 
to account for the band-gap problem in optical excitation.
%
For extended systems a long-range correction has been proposed \cite{botti2004long} 
to approximately account for the inherent non-locality of screening,
which however considerable increases the cost of computation.
The most accurate estimation of linear optical excitation is possible through solution of the Bethe-Salpeter equation(BSE) 
 \cite{strinati1982dynamical} 
in the basis of exciton amplitudes considered as products of KS single particle levels from the valence and conduction bands.
The high computational cost of BSE calculation stems from convergence of the interaction kernel and the exciton amplitude with respect unoccupied
states. 
%
%
%
In particular for the large nano-diamond systems considered in this work, DFT+GW and subsequent BSE calculations become 
computationally impossibly expensive.
%

In this work we propose a computationally much inexpensive alternate route to estimate optical absorption spectra, 
primarily the threshold up to workable accuracy, using
the real time dynamics(RTD) approach, in a minimal tight-binding basis constructed from first principles.
The minimal nature of the basis stems from the fact that the optimally directed hybrid orbitals
 \cite{de2023maximally}, 
maximally incorporate all covalent interactions
prevalent in the system and thus represent nearest neighbor covalent bonds predominantly by a single off-diagonal element.
However given the ideal nature of the bond angles in the systems consider in this work
the degenerate hybrid orbitals \cite{hossain2021hybrid} constitute the minimal basis.
We have recently demonstrated effective transfer of self-energy correction from smaller reference system to much larger
isomorphic systems in the basis of such  directed  hybrid  atomic Wannier orbitals (HAWO),
to render estimates for quasi-particle band-gap     \cite{hossain2020transferability,hossain2021hybrid} of the large systems
with over 90\% accuracy. 
In this work we use self-energy corrected tight-binding parameters in the HAWO basis and introduce a justifiably
parametrizable anti-causal 
correction to RTD which indirectly accounts for the presence of excitons in the Hamiltonian
to render  absorption threshold of large systems with workable accuracy comparable to GW-BSE estimates. 
%
\begin{figure}[b]
\includegraphics[width=8cm]{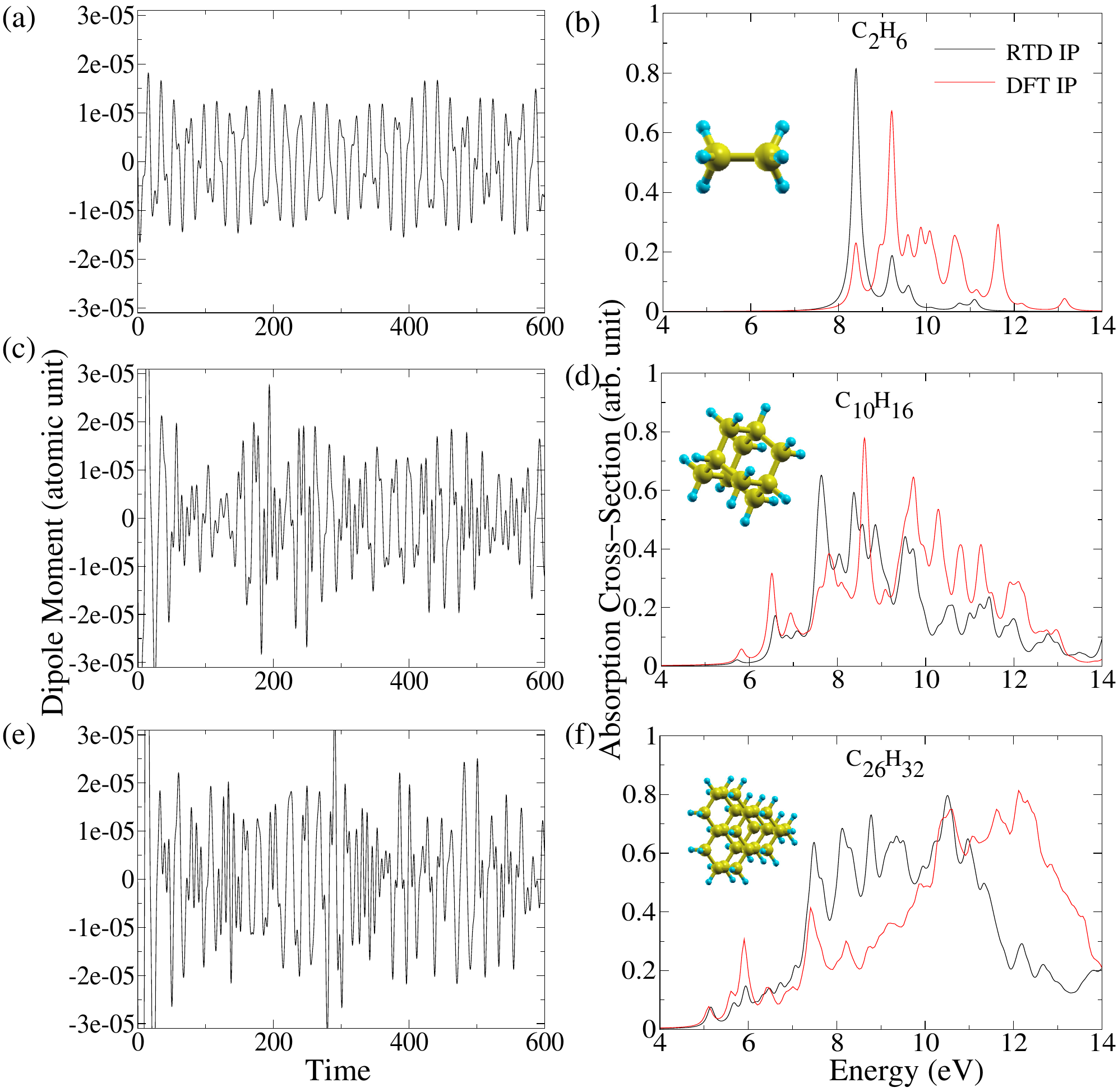}
\caption{RTD(IP) polarization in the directed TB basis without anti-causal correction, for 
:(a) C$_2$H$_6$, (c) Adamantane, (e) Pentamantane.
The corresponding RTD spectra for longer time, and the spectra obtained from the LR density-density response function
computed from the KS single-particle states, for he same systems (b,d,f respectively).}
\label{indep}
\end{figure}
\begin{figure}[t]
\includegraphics[width=8cm]{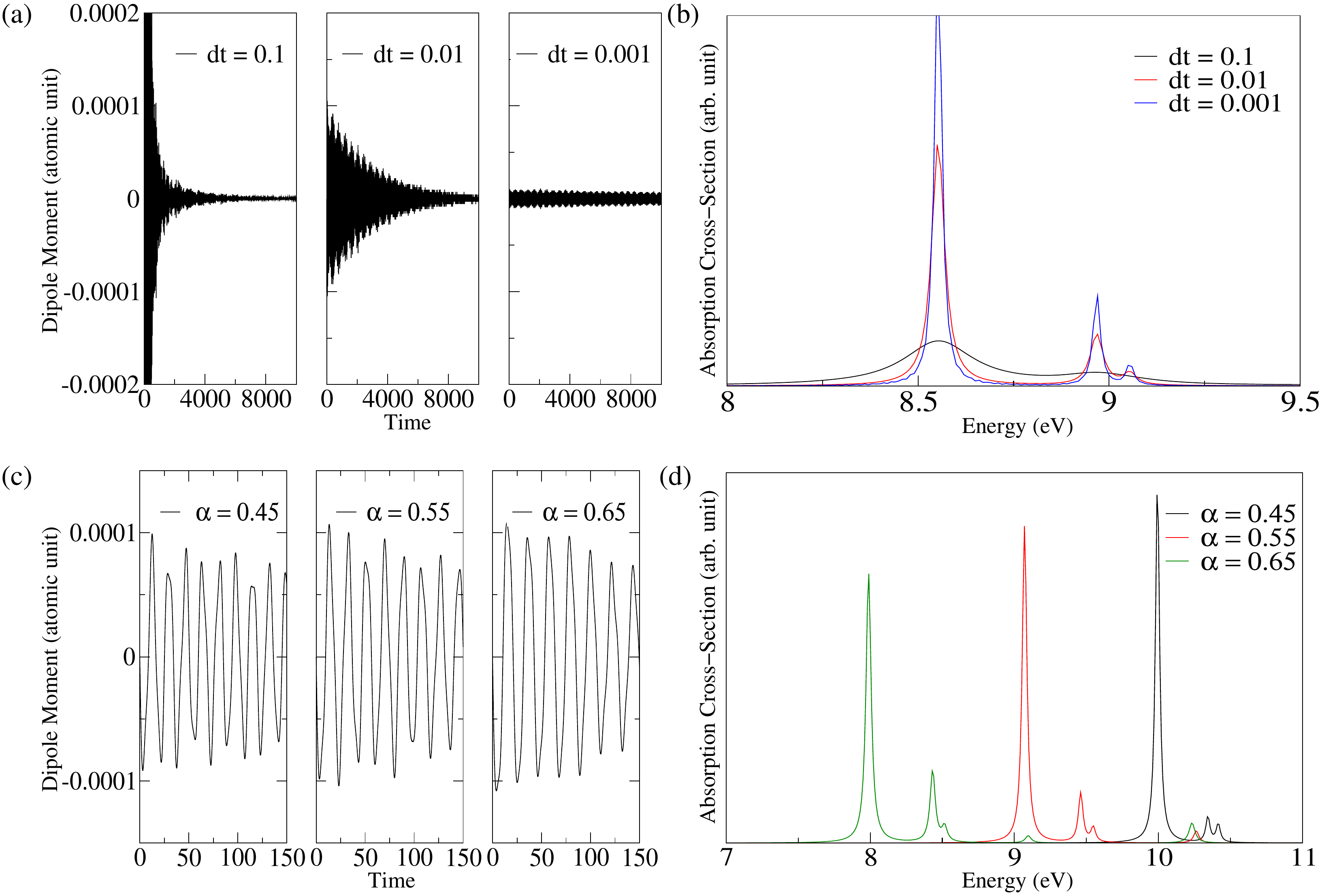}
\caption{Nature of anti-causally corrected polarization(a,c) and absorption spectra(b,d) for: (a,b) different time step 
$dt$ with $\alpha=0.6$, and (c,d) for different values of $\alpha$ with same time step $dt=0.01$a.u..}
\label{pol}
\end{figure}
\section{Methodology}
\label{method}
First we briefly sketch the construction of the tight-binding Hamiltonian from first principles in the basis of HAWOs \cite{hossain2021hybrid}, which are constructed from the KS states of the given system as:
%
\begin{equation}
\Phi_{j}(\vec{r})=\sum_{\vec{k}} e^{i \vec{k} \cdot\vec{r}}\sum_{l}U_{jl} \Psi^{KS}_{\vec{k},l}(\vec{r})
	\label{hawo}
\end{equation}
%
where the choice of gauge (U) is obtain though Lowdin symmetric orthogonalization of the template made of hybrid atomic orbitals (HAO) projected on the KS states.
The four $sp^3$ HAOs of C or Si atoms are obtained through maximal joint diagonalization  \cite{cardoso1996jacobi}
of the first moment matrices corresponding to the three orthogonal direction in the basis of the lowest four KS states generated for an isolated  atom using the same pseudo potential used for the nano-diamond systems.
%
%
HAWOs being directed toward nearest neighbor coordination, TB parameters derived in such a basis are therefore easily transferable from smaller to larger nano-diamond systems through mapping of neighbourhood.
Self-energy corrected TB parameters for samller reference systems are obtained in the HAWO basis 
using quasi-particle levels computed at the $G_0W_0$ level \cite{hedin1965new,hedin1969advances}. 

For time dynamics we use the standard Crank-Nicholson \cite{varga1962matrix} 
approach where the time evolution operator is approximated as:
\begin{equation}
 {\hat U}(t+\delta t,t)
 = \frac{1-i{\hat H}(\delta t/2)}{1+i {\hat{H}}(\delta t/2)}
\label{eqn:cls}
\end{equation}
which is exact up to $\delta t^2$ and assumes \ham\ to remain constant over the time interval $\delta t$.
In our implementation we have used a higher order formulation \cite{varga1962matrix}
of ${\hat U}(t+\delta t,t)$:
\begin{equation}
        {\hat U}(t+\delta t,t)=\frac{\big(I-\frac{i\delta t}{2\hbar}H-1/2(H\delta t/2)^2+i1/6(H\delta t/2)^3\big)}{\big(I+\frac{i\delta t}{2\hbar}H-1/2(H\delta t/2)^2-i1/2(H\delta t/2)^3\big)}
\end{equation}

%
We apply a small constant electric field across the system \cite{yabana2006real} $\vec{E}(\vec r,t)=\vec{I}\delta(t)$
in the first time-step and revert back to the original unperturbed 
\ham\ from the second time-step onwards.
If the \ham\ is retained unchanged for all successive steps then 
the absorption spectra obtained from the time evolved occupied states can be compared to the spectra obtained at 
the independent particle level within linear response, as demonstrated in Fig.\ref{indep}. 
%

Photo-absorption cross-section relevant to the direction $\mu$ is calculated as:
\begin{equation}
 \sigma_{\mu\mu}(\omega)=\frac{4\pi\omega}{c}\Im[\alpha_{\mu\mu}(\omega)]
\end{equation}
where the polarizability $\alpha_{\mu\mu}(\omega)$ is calculated from the
$\mu$-th component of the time evolved net dipole-moment $\vec{p}(t)$, as:
\begin{equation}
 \alpha_{\mu\mu}(\omega)=\frac{1}{\tilde{E}_\mu(\omega)}\int dt e^{i\omega t}p_{\mu}(t),
\end{equation}
$\tilde{E}_\mu(\omega)$ being the Fourier transform of the applied electric field component $E_\mu(\vec r,t)$.
In effect $\tilde{E}_{\mu}$ is constant in  $\omega$ given the pulsed nature of the applied field.
The net dipole moment from the time  evolved occupied states is calculated approximately as:
\begin{equation}
\vec{p}(t)=-e \sum^{occ}_i \sum_j \mid C_{ij}(t)\mid^2 \vec{r}_j  
\end{equation}
$\left\{ \vec{r}_j \right\}$ being the charge centres of the directed hybrid orbital basis ${\left\{\phi_j \right\}}$,
and the time evolved states in effect being $\left\{ \psi_i(r,t)=\sum_j C_{ij}(t) \phi_j(r)  \right\}$.
%
%
Such a simultaneity is prevented by the inherent non-commuting nature
of more than one orthogonal position operators if projected within a finite subspace.
%
%
In Fig.\ref{indep}, we find good agreement of threshold and initial peaks between the 
IP level spectra obtained within linear response from the KS states and the RTD spectra rendered by the 
time evolved occupied states in the HAWO basis.
%

In order to compute spectra beyond the independent particle level we invoke an anti-causal correction, 
where Hamiltonian is updated at each time step 
as per the evolution of the charge density in the previous step.
%
A correction $\{\delta V_j\}$ is added to the on-site terms of the unperturbed Hamiltonian 
$H_0$ such that the eigenstate of the modified Hamiltonian renders to charge density of the
previous time step.
The applied correction is anti-causal since it
is derived back from $\{\delta \rho_j\}(t-\delta t)$.
%
%
The \hamt\ is thus updated such that the instantaneous charge density will become increasingly similar to that of the previous step as time progresses, as we see in Fig.\ref{pol}(a).  
%
%
%
In our implementation we compare the evolving charge density at each time step to that of the ground state and use density-density response function or the dielectric susceptibility $(\chi)$ computed from the ground state wave functions in the HAWO basis.

At the outset, $\delta V(\vec{r})$ can be connected to $\delta \rho(\vec{r'})$ through \kaiinv\ as 
\[
\delta V(\vec{r}) = \int d \vec{r'} \chi^{-1}(\vec{r},\vec{r'}) \delta{\rho}(\vec{r'}).
\]
With $\delta \rho(\vec{r'})=\sum_j \delta \rho_j(\vec{r'})$, $j$ being orbital index.
We can notionally discretise $\chi^{-1}(\vec{r},\vec{r'})$ as $\chi_{ij}^{-1}\Theta(\vec{r}-\vec{r}_i)\Theta(\vec{r'}-\vec{r}_j)$  
using non-intersecting step-functions $(\Theta)$ centered at orbital charge centers $\{r_i\}$, leading to :
%
\begin{eqnarray}
& &\delta V_i(\vec{r}) 
=  \int d \vec{r'} \sum_j [\chi_{ij}^{-1}\Theta(\vec{r}-\vec{r}_i)\Theta(\vec{r'}-\vec{r}_j)]
\delta{\rho}_j(\vec{r'}) \nonumber \\
	& = &\Theta(\vec{r}-\vec{r}_i) \sum_j\chi_{ij}^{-1} \int d \vec{r'}  [\Theta(\vec{r'}-\vec{r}_j)]
\delta{\rho}_j(\vec{r'}),
\label{dvone}
\end{eqnarray}
which implies
\begin{equation}
	\delta V_i=\sum_j \chi_{ij}^{-1}{\rho}_j,
\end{equation}
where 
$\delta V_i(\vec{r})=\delta V_i\Theta(\vec{r}-\vec{r}_i)$ and $\rho_j=\int d \vec{r'}  [\Theta(\vec{r'}-\vec{r}_j)]
\delta{\rho}_j(\vec{r'})$. 
In our case we obtain:
\begin{equation}
 \delta \rho_j(t)=\sum^{occ}_i (\mid C_{ij}(t)\mid^2 - \mid C^0_{ij}\mid^2,
\label{drho}
\end{equation}
where $\left\{C_{ij}\right\}$  are the elements of the $i$-th wave function, which are also used to estimate 
\kai\ in the static limit within linear response as:
%
\begin{equation}
	\chi(i,j)=\sum^N_{kl} (f^0_k - f^0_l)\frac{C^0_{li}C^{0*}_{ki}C^{0*}_{lj}C^0_{kj}}{E^0_k- E^0_j + i \eta },
\label{chi}
\end{equation}
where $f^0_i$ and $E^0_i$ respectively are the occupation and energy of the $i$-th level in the ground state,
and $\eta$ is a broadening parameter. 

Although partitioning the unit-cell into such non-intersecting domains around the centers of each orbitals should in-principle possible,
it would also results into sharp changes in values of $\chi^{-1}$ across neighboring domains, which is unrealistic. 
%
Instead, we can relax the criterion of non-intersection and let spherical domains defined by a cut-off radius around the charge centers to enclose substantial fractions of the 
respective orbitals, and introduce corrections to avoid multi-valuedness of contributions to $\delta V_i$
due to overlapping orbitals.
%
%
The unit cell can then be partitioned into different regions of different numbers of overlapping orbitals.
The total charge of an orbital can be partitioned into such regions as:
\begin{equation}
	\rho_i=\rho_{ii}+\sum_{j\ne i}\rho_{ij}+\sum_{(j<k)\ne i}\rho_{ijk}+\sum_{(j<k<l)\ne i}\rho_{ijkl}+\dots
\end{equation}
where $\rho_{ii}$ is the part of $\rho_i$ which has no overlap with any other orbital, 
while $\{\rho_{ij,i\ne j}\}, \{\rho_{ijk,(j<k)\ne i}\}$ and $ \{\rho_{ijkl,(j<k<l)\ne i}\}$ are
parts of $\rho_i$ in regions with 2, 3 and 4 overlapping orbitals.
Similar partitioning can be done for $\delta \rho_i$ where contributions can be approximately obtained, for example, as:
\begin{equation}
\delta\rho_{ijk}=\delta\rho_i\frac{\rho_{ijk}}{\rho_i}.
\label{drhopart}
\end{equation}
Values of the contributions would thus depend on the spherical volume considered for each of the orbitals.
In this work we find that it is appropriate to consider a spherical region around the charge centers of each orbital such that
the sphere encloses about 60\% of the total normalization of each orbital.
%
%
Adding the contribution from the different regions, we can write: 
\begin{equation}
        \delta V_I =  \sum_{j=1}^{N}\left\{\delta V_{Ij}+\sum_{k>j}^{N}\left\{\delta V_{Ijk}+\sum_{l>k>j}^{N}\left\{\delta V_{Ijkl}+\dots\right\}\right\}\right\}\nonumber
\end{equation}
where we have followed the convention of denoting the overlap regions in an ascending order of indexes to avoid over counting. 
Considering average contributions from each of the regions of overlapping orbitals, we can write as:
\begin{eqnarray}
	&&\delta V_I=\sum_{j=1}^{N}\chi_{Ij}^{-1}\delta\rho_{jj} + \sum_{j=1}^N\sum_{k>j}^{N}\frac{1}{2}\left(\chi_{Ij}^{-1}\delta\rho_{jk}+\chi_{Ik}^{-1}\delta\rho_{kj}\right)\nonumber\\
		  &+& \sum_{j=1}^N\sum^N_{k>j}\sum_{l<k<j}^{N}\frac{1}{3}\left(\chi_{Ij}^{-1}\delta\rho_{jkl}+\chi_{Ik}^{-1}\delta\rho_{klj}+\chi_{Il}^{-1}\delta\rho_{ljk}\right)+\dots\nonumber\\
	&&=\sum_{j=1}^{N}\chi_{Ij}^{-1}\left(\frac{\rho_{jj}}{\rho_j}+\frac{1}{2}\sum_{j<k}\frac{\rho_{jk}}{\rho_j}+\frac{1}{3}\sum_{j<k<l}\frac{\rho_{jkl}}{\rho_j}+\dots\right)\delta\rho_j\nonumber
\end{eqnarray}
using Eqn.(\ref{drhopart}), implying:
\begin{equation}
 \delta V_I=\sum_{j=1}^{N}\chi_{Ij}^{-1}\alpha_j\delta\rho_j
\end{equation}
where
\begin{equation}
\alpha_j=\left(\frac{\delta \rho_{jj}}{\rho_j}+\frac{1}{2}\sum_{j<k}\frac{\rho_{jk}}{\rho_j}+\frac{1}{3}\sum\frac{\rho_{jkl}}{\rho_j}+\dots\right)
\label{alpha}
\end{equation}
\begin{figure}[b]
\includegraphics[width=8cm]{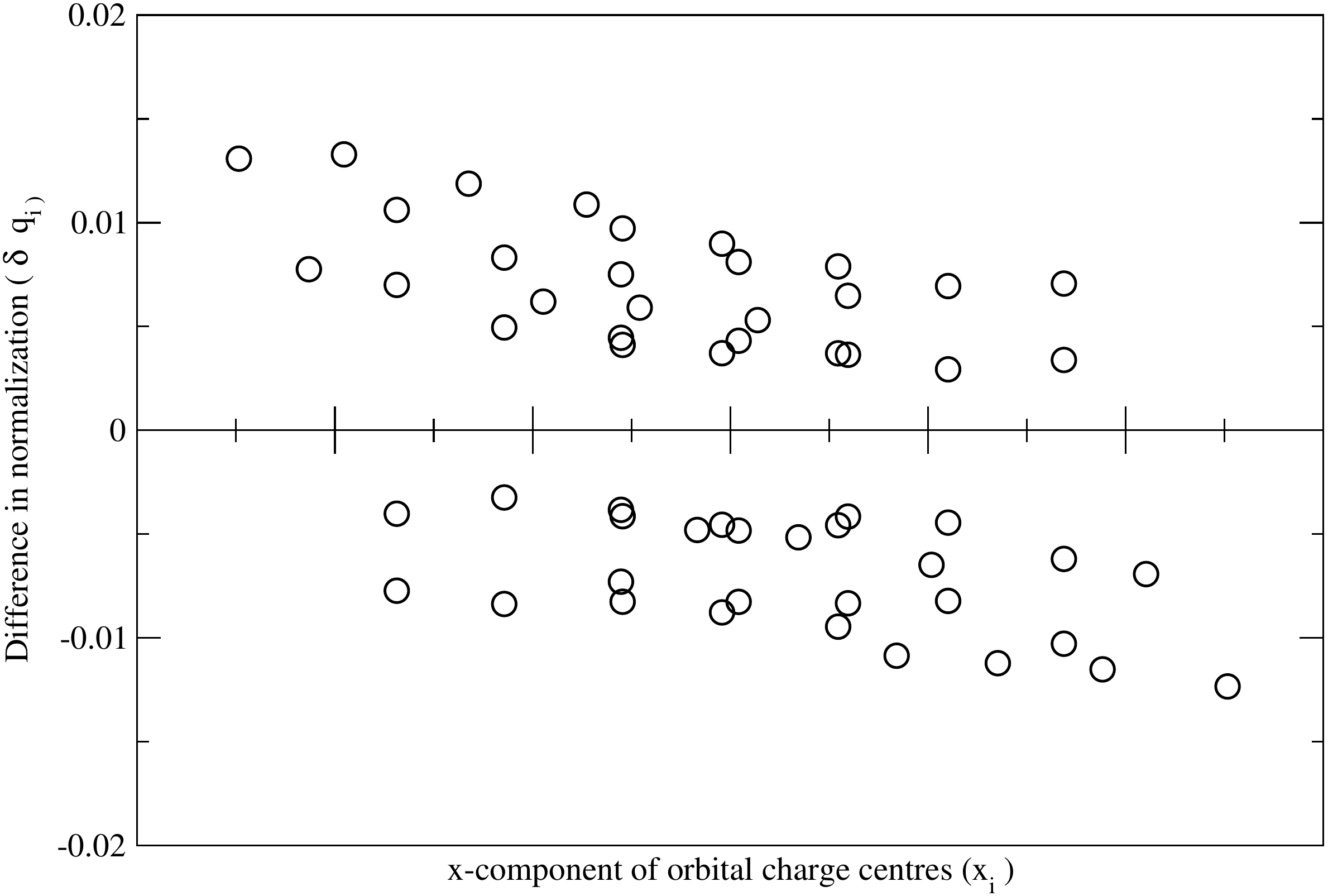}
\caption{Change in charge $q_i$ associated with orbitals due to a static electric field applied to the TB Hamiltonian 
of adamantane.}
\label{ada}
\end{figure}
\begin{figure}[t]
\includegraphics[width=8cm]{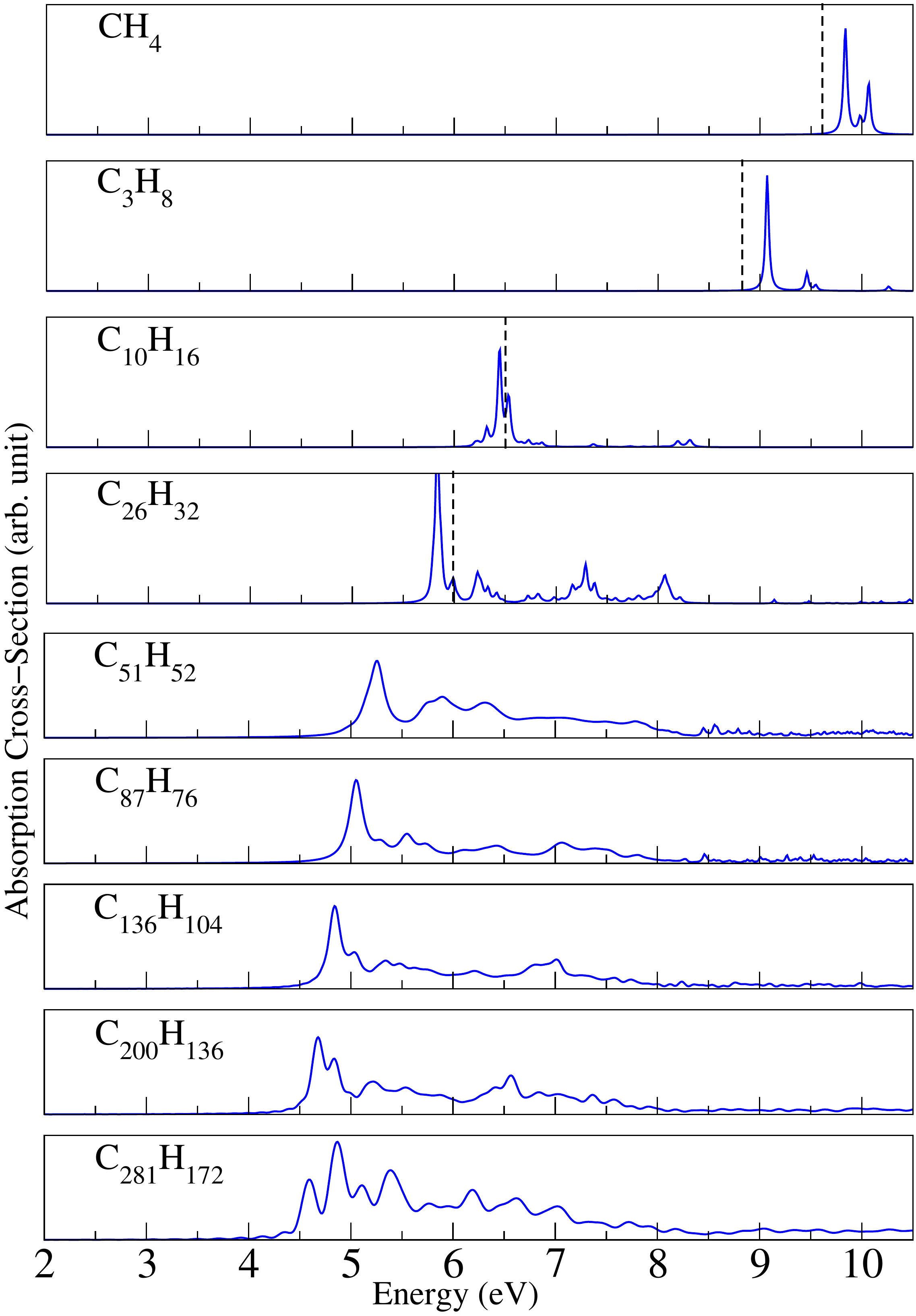}
\caption{ACC-RTD based optical absorption spectra of H passivated C clusters and nano-diamonds for $\alpha=\bar\alpha=0.55$
(used up to propane) and $\alpha=\alpha_H=0.75$ (used for adamantane and beyond)
corresponding to the enclosed normalization of 0.6. The dashed lines are experimental absortion thresholds
 \cite{koch1971optical,landt2009optical}.}
\label{carbon1}
\end{figure}
\begin{figure}[b]
\includegraphics[width=8cm]{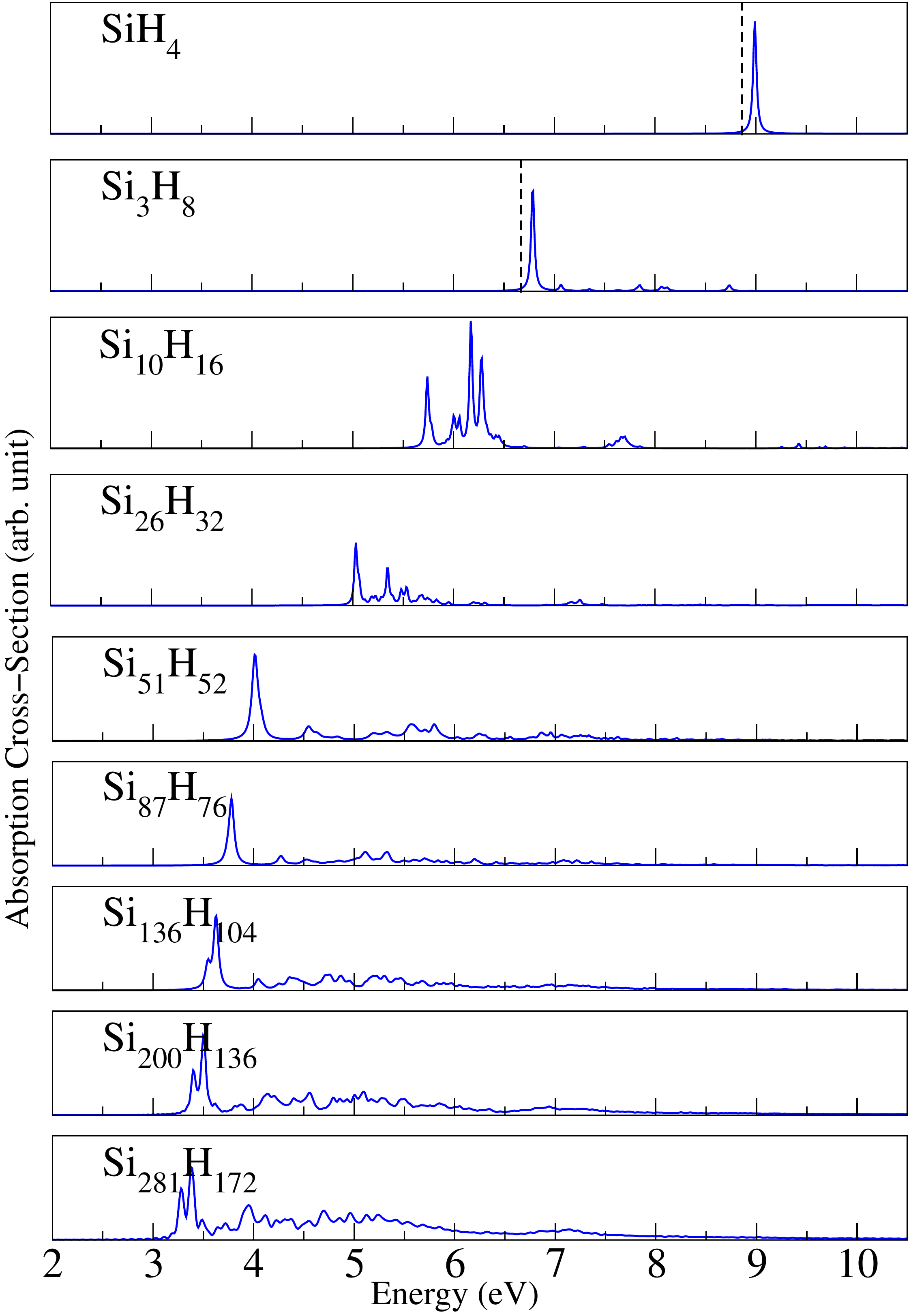}
\caption{ACC-RTD based optical absorption spectra of silicon clusters and nano-diamonds for $\alpha=\bar\alpha=0.55$ (used for SiH$_4$)
and $\alpha=\alpha_H=0.65$ (used for Si$_3$H$_8$ onwards)
corresponding to the enclosed normalization of 0.6. The dashed lines are lowest experimental absorption peaks approximately
estimated from literature \cite{itoh1986vacuum,tiago2006optical} }. 
\label{silicon1}
\end{figure}

%
The values of $\{\alpha_j\}$ are therefore a function of cut-off radius to be defined for the orbitals. The role of $\{\alpha_j\}$ is to rescale the contribution $\chi_{Ij}^{-1}\delta\rho_j$ to $\delta V_I$ due to overlap of $\delta \rho_j$ with
that of other orbitals $\delta\rho_{i\ne j}$.

\section{Computational detalis}
\label{compute}
Ground state electronics structure of all the systems considered were calculated using the Quantum
Espresso (QE) code \cite{giannozzi2009quantum}, which is a plane wave based implementation of density functional theory (DFT) \cite{hohenberg1964inhomogeneous,kohn1965self}.
 We used norm-conserving pseudo-potentials with the Perdew-Zunger (LDA) exchange-correlation \cite{perdew1981self} functional 
 and a kinetic energy cut-off of 80 Ry for plane wave basis and four time more for charge density.
Self-energy corrections to the single particle levels have been estimated at the non-self-consistent single-shot $G_0W_0$ level 
 of the GW approximation implemented in the BerkeleyGW code \cite{deslippe2012berkeleygw}.
 %
 Generalized plasmon-pole (GPP) model \cite{hybertsen1986electron} 
 was used to extend the static dielectric function to the ﬁnite frequencies. 
 Quasi-particle energies were converged for pentamantane using excess of 4000 empty states.
%
In-house implementations were used to construct HAOs and HAWOs respectively from the KS states of isolated atoms and the given systems. 
%
TB parameters in the HAWO basis were obtained using the self-energy corrected single particle levels.
 Self energy corrected TB parameters were transferred from reference to target systems for nano-diamonds larger than pentamantane through mapping of neighborhood beyond nearest neighbors. 
The IP level absorption spectra from the density-density response function has been 
calculated using the Yambo code \cite{marini2009yambo}. 
\begin{figure}[t]
\includegraphics[width=8cm]{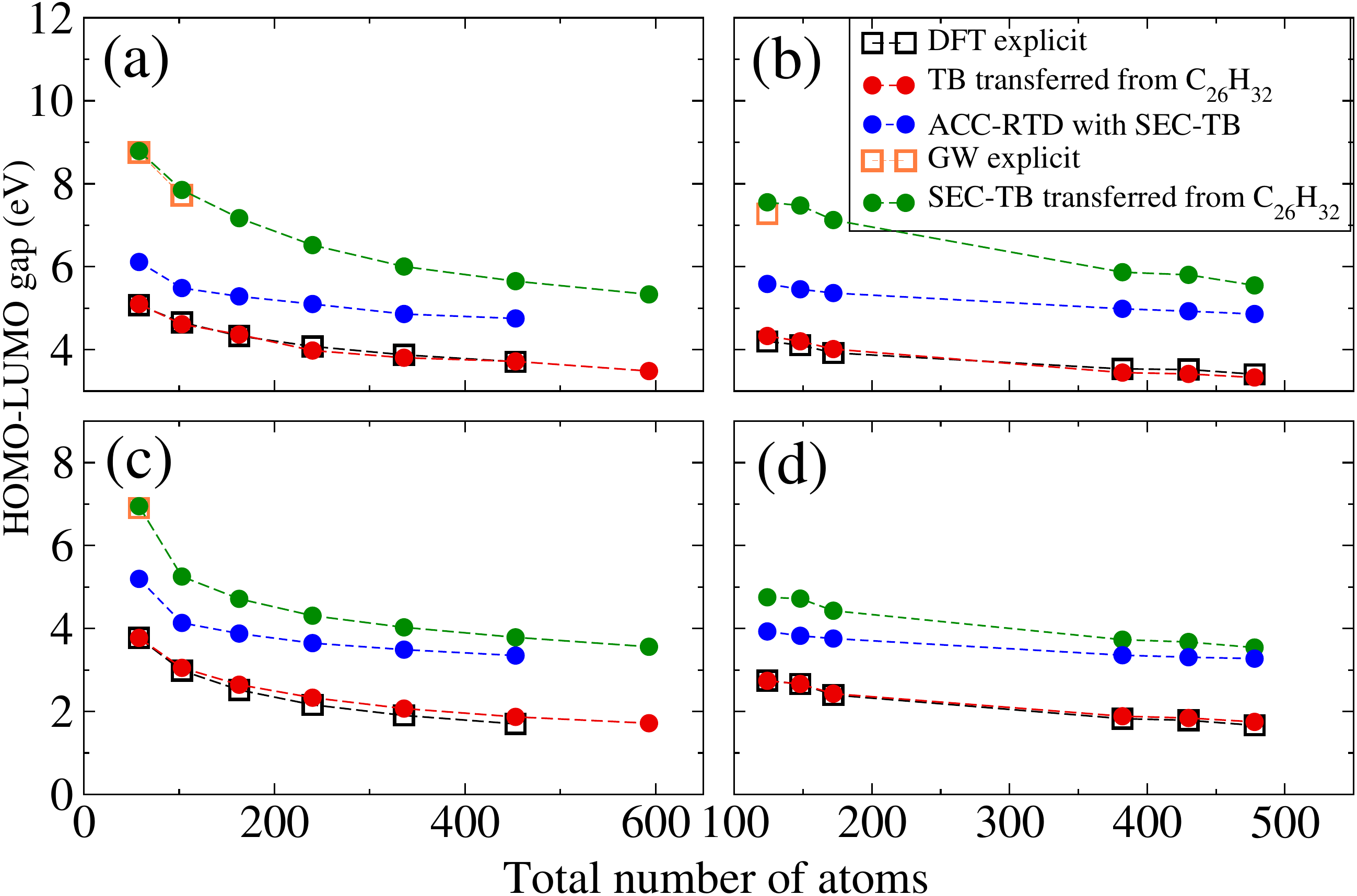}
\caption{Comparison of optical absorption threshold obtained using ACC-RTD and estimates of HOMO-LOMO gaps
with and without self-energy correction, for pyramidal (a,c) and bi-pyramidal (b,d) systems made of
C (a,b) and Si (c,d).}
\label{gapsummary}
\end{figure}
\section{Results and Discussion}
To test the efficacy of the RTD approach at the IP level in the directed tight-binding basis within linear response, 
we  compare the spectra obtained using RTD without any anti-causal correction
to the spectra computed from the density-density response function 
evaluated at the IP level \cite{marini2009yambo} 
using exactly the same KS states used to compute the TB parameters. 
%
As suggested by Fig.\ref{indep}(a,c,e) polarization fluctuation grows more non-uniform with increasing size of nano-diamonds
 due different contributions from different regions like vertices, surfaces and ridges. 
%
The agreement of peak positions in two spectra [Fig.\ref{indep}(b,d,f)], particularly near the threshold,  
suggests nominal loss of spectroscopic information due to representation of polarization in terms of charge centers of the
HAWO basis orbitals. 

In Fig.\ref{pol}, we assess the effect of anti-causal correction, and duration of time step ($\delta t$), on the evolution of 
polarization and the consequent absorption spectra.
The rate of decay in amplitude of polarization decreases with decreasing 
$\delta t$[Fig.\ref{pol}(a)],
implying expectedly,
lesser perturbation at each step, and thereby lesser mixing of states, leading to their increased life time, causing sharper absorption spectra[Fig.\ref{pol}(b)].
Whereas, the red shift of the calculated absorption threshold with increasing $\alpha$ 
is suggestive of less frequent polarization fluctuation as a result of the anti-causal correction,
which is evident in Fig.\ref{pol}(c) as we see lesser number of polarization oscillation within same real time as $\alpha$ increases.
Estimation of $\alpha$ is thus crucial for accuracy of the absorption threshold rendered by the anti-causally corrected RTD (ACC-RTD) 
proposed in this work.
As evident from the derivation of Eqn.\ref{alpha} the parameter $\alpha$ is
essentially a correction factor to approximately account for over counting of charge density due to overlapping of orbitals while associating net charges to orbital charge centers. 
The approximation is exclusively based on a cutoff radius chosen to define a spherical region around the charge center of each orbital.
In fact, instead of fixing the cutoff radius, we find it convenient to fix the normalization enclosed by the spherical region 
to determine the cutoff radius. 
Notably, we need not restrict hopping parameters based on this cutoff radius in our calculation, since we use it only
to derive correction factors to account for dominant overlap of orbitals.
%
%
For both the C and Si based nano-diamonds we find normalization of about 0.6,
implying cut-off radius of about 0.8\AA\, 1.22\AA\ and 0.92\AA\ for orbitals of
C, Si and H respectively, to render values of $\alpha$ which match the observed absorption thresholds,
well within an error of 0.5 eV. 
Similar values of enclosed normalization should therefore work for molecules and clusters with covalent bonds
made of $sp^3$ orbitals with principle quantum number starting with 2 in general.
Notable that these radius cutoffs are close to the the covalent radii of C and Si in nano-diamonds and bulks.
However, for stability of the RTD evolution, we need to use a constant $\alpha$ since the unperturbed Hamiltonian
does not have any inhomogeneity other than the on-site terms.
The choice of constant $\alpha$ should depend on the
orbitals which are the dominant contributors to the net polarization. 
With growing system size the key contributions to polarization increasingly arise from peripheral atoms due to cancellation 
of contributions from neighboring atoms in the interior of the systems.
This is evident in Fig.\ref{ada}, where we have plotted
contribution to polarization from different orbitals across adamantane due to the static
electric field applied in the TB Hamiltonian. 
%
Accordingly, with increasing system size, from methane to adamantane ($C_{10}H_{16}$), we find that the  $\alpha$ required 
to match the experimentally observed\cite{koch1971optical,landt2009optical} absorption threshold, or the GW+BSE based estimation
\cite{demjan2014electronic,yin2014quasiparticle}
of the onset of adsorption, 
 evolves from being closer to the average value ($\bar\alpha=\sum^N_i\alpha_i/N$) to 
that of the peripheral $1s$ orbitals of H ($\alpha_H$), as apparent in Fig.\ref{carbon1},
where the values of $\alpha$ used for methane and propane are that of  $\bar\alpha$,
whereas adamatane onwards we have used $\alpha_H$.
%
%
%
%
%
In-fact, for Si nano-clusters and nano-diamonds[Fig.\ref{silicon1}], which are larger in size than their C based counterparts, 
$\alpha_H$ renders a satisfactory match with the observed absorption threshold 
\cite{itoh1986vacuum}
or the same estimated from GW+BSE
\cite{tiago2006optical},
 for Si$_3$H$_8$ itself.
It is thus  expected that for both, C and Si based larger nano-diamonds, 
with $\alpha_H$ as per a radius cutoff enclosing about 0.6 normalization,  
ACC-RTD would render absorption threshold well within an error of 0.5eV,  
with self-energy corrected TB parameters transferred from pentamantane. 
%
%
In Fig.\ref{gapsummary} we summarize the comparison of the absorption threshold estimated using the ACC-RTD approach 
for larger nano-diamonds starting with pentamantane, an HOMO-LUMO gaps computed explicitly within DFT as well as DFT+G$_0$W$_0$,
and the same estimated using TB parameters transferred from pentamantane with and without self-energy correction.
%
\section{Conclusion}
In this work we present an anti-causally corrected real time dynamics (ACC-RTD) approach within a self-energy corrected 
minimal tight-binding(TB) framework for computationally inexpensive  estimation of primarily the optical absorption threshold 
for large systems with hundreds of atoms.
The  anti-causal correction based on the density-density response function computed inexpensively in the minimal TB basis constitutes the key mechanism
incorporating the effect of transfer of electron from valence to conduction band. 
The minimal multi-orbital TB basis is constituted by the hybrid atomic Wannier orbitals(HAWO) 
which are constructed from the Kohn-Sham single particle states of the system and are 
directed towards the nearest neighbor coordination by construction.
Self-energy corrected TB (SEC-TB) parameters in the HAWO basis are computed within the DFT+G$_0$W$_0$ framework 
for smaller clusters and transferred to large nano-diamonds through mapping of neighborhood.
Thus for large systems, with transferred SEC-TB parameters, ACC-RTD parametrized as per the major contributors to dipole-moment,
can be expected to render absorption thresholds 
comparable with  GW-BSE based estimates with a small fraction of computational resource required for explicit computation.
%
\section{Acknowledgements}
Computations have been performed in computing clusters supported by the 
Dept. of Atomic Energy of the Govt. of India.
MH acknowledges the support from of Ministry of Education Centre of Excellence for Novel Energy Material (CENEMA: RP-074) during his stay in NISER. 

\end{document}